\documentclass[prl,preprint,showpacs,superscriptaddress,
tightenlines,byrevtex]{revtex4}

\usepackage{graphicx}
\usepackage{dcolumn}

\begin{document}

\preprint{TRI-PP-01-11}

\title{Parity Violation in Proton-Proton Scattering at 221 MeV}

\author{A.~R.~Berdoz}
\author{J.~Birchall}
\author{J.~B.~Bland}
\affiliation{Department of Physics and Astronomy, University of Manitoba,
Winnipeg, MB, Canada R3T 2N2}
\author{J.~D.~Bowman}
\affiliation{Physics Division, Los Alamos National Laboratory,
Los Alamos, NM 87545, USA}
\author{J.~R.~Campbell}
\affiliation{Department of Physics and Astronomy, University of Manitoba,
Winnipeg, MB, Canada R3T 2N2}
\author{G.~H.~Coombes}
\affiliation{TRIUMF, 4004 Wesbrook Mall, Vancouver, BC, Canada V6T 2A3}
\author{C.~A.~Davis}
\affiliation{Department of Physics and Astronomy, University of Manitoba,
Winnipeg, MB, Canada R3T 2N2}
\affiliation{TRIUMF, 4004 Wesbrook Mall, Vancouver, BC, Canada V6T 2A3}
\author{A.~A.~Green}
\affiliation{Department of Physics and Astronomy, University of Manitoba,
Winnipeg, MB, Canada R3T 2N2}
\author{P.~W.~Green}
\affiliation{Centre for Subatomic Research, University of Alberta,
Edmonton, AB, Canada T6G 2N5}
\author{A.~A.~Hamian}
\affiliation{Department of Physics and Astronomy, University of Manitoba,
Winnipeg, MB, Canada R3T 2N2}
\author{R.~Helmer}
\author{S.~Kadantsev}
\affiliation{TRIUMF, 4004 Wesbrook Mall, Vancouver, BC, Canada V6T 2A3}
\author{Y.~Kuznetsov}
\thanks{deceased}
\affiliation{Institute for Nuclear Research, Academy of Sciences of Russia,
RU-117334 Moscow, Russia}
\author{L.~Lee}
\affiliation{Department of Physics and Astronomy, University of Manitoba,
Winnipeg, MB, Canada R3T 2N2}
\author{C.~D.~P.~Levy}
\affiliation{TRIUMF, 4004 Wesbrook Mall, Vancouver, BC, Canada V6T 2A3}
\author{R.~E.~Mischke}
\affiliation{Physics Division, Los Alamos National Laboratory,
Los Alamos, NM 87545, USA}
\author{S.~A.~Page}
\author{W.~D.~Ramsay}
\author{S.~D.~Reitzner}
\affiliation{Department of Physics and Astronomy, University of Manitoba,
Winnipeg, MB, Canada R3T 2N2}
\author{T.~Ries}
\affiliation{TRIUMF, 4004 Wesbrook Mall, Vancouver, BC, Canada V6T 2A3}
\author{G.~Roy}
\author{A.~M.~Sekulovich}
\affiliation{Department of Physics and Astronomy, University of Manitoba,
Winnipeg, MB, Canada R3T 2N2}
\author{J.~Soukup}
\author{G.~M.~Stinson}
\author{T.J.~Stocki}
\affiliation{Centre for Subatomic Research, University of Alberta,
Edmonton, AB, Canada T6G 2N5}
\author{V.~Sum}
\affiliation{Department of Physics and Astronomy, University of Manitoba,
Winnipeg, MB, Canada R3T 2N2}
\author{N.~A.~Titov}
\affiliation{Institute for Nuclear Research, Academy of Sciences of Russia,
RU-117334 Moscow, Russia}
\author{W.~T.~H.~van Oers}
\author{R.~J.~Woo}
\affiliation{Department of Physics and Astronomy, University of Manitoba,
Winnipeg, MB, Canada R3T 2N2}
\author{S.~Zadorozny}
\author{A.~N.~Zelenski}
\affiliation{Institute for Nuclear Research, Academy of Sciences of Russia,
RU-117334 Moscow, Russia}

\collaboration{The TRIUMF E497 Collaboration}
\noaffiliation

\date{\today}

\begin{abstract}

The parity-violating longitudinal analyzing power, $A_z$, has been measured
in $\vec{p}p$ elastic scattering at an incident proton energy of 221 MeV.
The result obtained is $A_z=(0.84 \pm 0.29 (stat.) \pm 0.17 (syst.)) \times
10^{-7}$. This experiment is unique in that it selects a single parity
violating transition amplitude $(^{3\!\!}P_2-^{1\!\!}D_2)$ and consequently
directly constrains the weak meson-nucleon coupling constant $h^{pp}_\rho$.
When this result is taken together with the existing $\vec{p}p$ parity
violation data, the weak meson-nucleon coupling constants $h^{pp}_\rho$ and
$h^{pp}_\omega$ can, for the first time, both be determined.

\end{abstract}


\maketitle

At the fundamental level, the weak interaction between nucleons is due to
heavy boson ($W^{\pm}$ and $Z^0$) exchanges between quarks. However, at low
and intermediate energies a single meson exchange model with one strong,
parity conserving, and one weak, parity non-conserving vertex is often used
to describe parity violating effects. The parameters of the strong vertex
are already quite well measured, while the weak interaction is
parameterized by a set of six weak meson-nucleon coupling constants
$h^1_\pi$, $h^{0,1,2}_\rho$, and $h^{0,1}_\omega$, where the superscripts
indicate the isospin change and the subscripts denote the exchanged meson.
These couplings were calculated by Desplanques, Donoghue, and Holstein
(DDH) \cite{DDH80} from the Weinberg-Salam model and quark bag wave
functions. Subsequent calculations of the weak meson-nucleon coupling
constants have been carried out by a number of authors
\cite{Dub86,Feld91,Kais89,Meis90,Meis99}, but it is apparent that the
constants still carry considerable ranges of uncertainty (see reviews
\cite{vano99,haeb95}).

The value of the weak meson-nucleon couplings has implications outside
nuclear parity violation. For example, Holstein\cite{Hols88} has concluded
that a small value of $h^1_\pi$ cannot be understood unless the current
algebra quark masses are increased by about a factor of two over the
original Weinberg values, which tends to produce a similar suppression of
theoretical estimates in other processes, e.g., the still questioned
$\Delta I = 1/2$ rule for flavour changing weak decays. In addition, more
precise values for the weak meson-nucleon couplings would permit better
theoretical calculation of the proton anapole moment
\cite{zhu00,Hast00,Haxt01} which, in turn, would improve calculations of
the axial vector radiative corrections needed to interpret electron-proton
parity violation experiments such as SAMPLE at MIT-Bates, and G-zero at
TJNAF.

Experimentally, the most accessible parity violating observable in $pp$
scattering is the longitudinal analyzing power, \mbox{$A_z = (\sigma^+ -
\sigma^-)/(\sigma^+ + \sigma^-)$}, where the $+$ and $-$ signs refer to the
helicity state (right or left handed) of the longitudinally polarized
incident proton beam, and $\sigma$ is the scattering cross section. This
letter reports the result for $A_z$ obtained by experiment 497 at TRIUMF.

A measurement of $A_z$ in p-p scattering is sensitive only to the short
range interaction mediated by $\rho$ and $\omega$ exchanges; the $\pi^0$ is
its own antiparticle, so parity violating $\pi^0$ exchange would also be CP
violating and is therefore suppressed by a factor of about $2 \times
10^{-3}$. Further, the TRIUMF 221 MeV measurement is designed to minimize
the effects of $\omega$ exchange. This is achieved by choosing the incident
energy so that the contribution to $A_z$ from the $(^1S_0 -^{3\!\!}P_0)$
transition amplitude integrates to zero over the acceptance of the
apparatus, leaving the contribution of the $(^3P_2 - ^{1\!\!}D_2)$
transition, which arises essentially only  from $\rho$
exchange\cite{Sim7581}.
Precise measurements of $A_z$ have already been made at
13.6 MeV \cite{Evers} and at 45 MeV \cite{Kist87}. However, unlike the
present measurement of $A_z$, which is proportional primarily to
$h^{pp}_\rho = h^0_\rho + h^1_\rho + h^2_\rho/\sqrt{6}$, the
low energy results are proportional to a linear combination of
$h^{pp}_\rho$ and $h^{pp}_\omega$, where $h^{pp}_\omega = h^0_\omega +
h^1_\omega$. By combining the present measurement with the low energy
result, $h^{pp}_\rho$ and $h^{pp}_\omega$ can be determined individually.

\begin{figure}
\includegraphics[width=\linewidth]{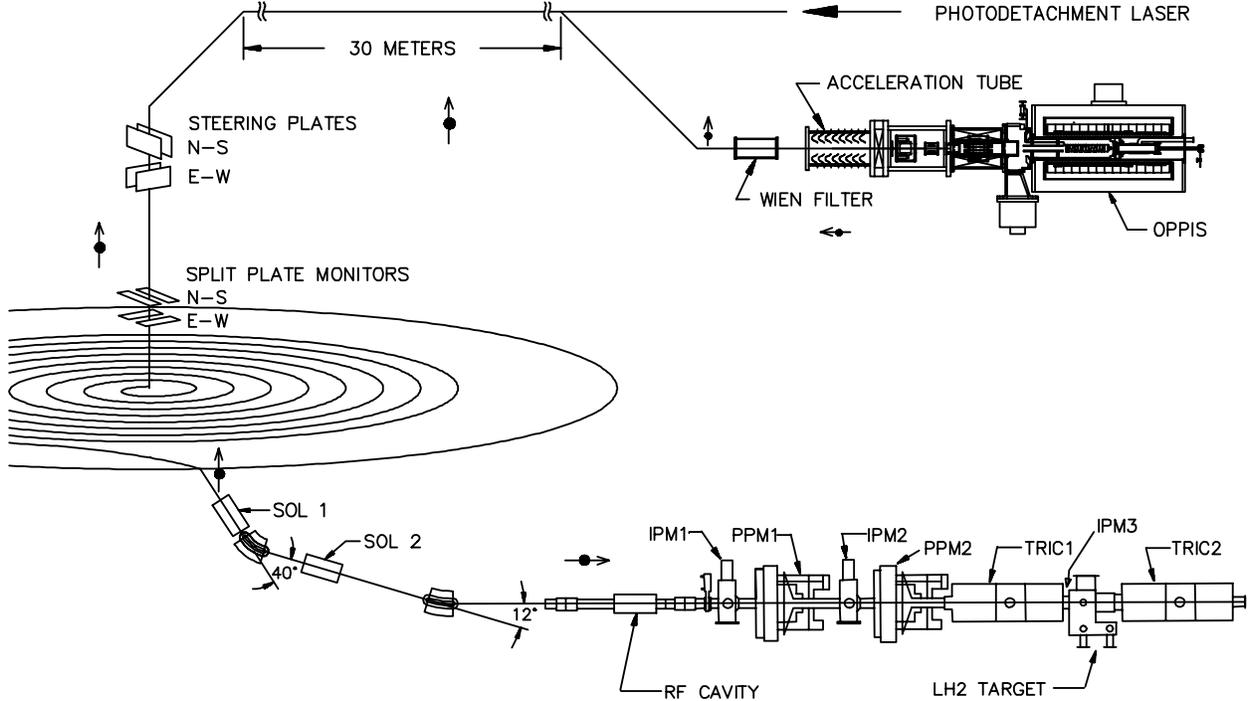}
\caption{General layout of the TRIUMF p-p parity violation
experiment.(OPPIS: Optically Pumped Polarized Ion Source; SOL: spin
precession SOLenoid magnet; IPM: Intensity Profile Monitor; PPM:
Polarization Profile Monitor; TRIC: TRansverse electric field Ionization
Chamber). One of eight possible spin configurations of ion source,
cyclotron and experiment is indicated.}
\label{layout}
\end{figure}

The layout of experiment 497 is shown in Figure \ref{layout}. A 200 nA
proton beam with a longitudinal polarization of $0.75\pm0.02$ is incident
on a 0.40 m long liquid hydrogen ($LH_2$) target. Transverse electric field
ion chambers (TRIC1 and TRIC2) located immediately upstream and downstream
of the target, measure the change in transmission when the helicity of the
incident beam is reversed.

The optically pumped polarized ion source (OPPIS)\cite{oppis} minimizes
unwanted helicity correlated modulations of beam properties. Helicity
reversals are implemented at OPPIS by changing the frequency of the lasers
which optically pump Rb vapor whose polarization is ultimately transferred
to the protons of the $H^-$ beam. No macroscopic electric or magnetic
fields are altered. The polarization of the Rb vapor is measured every spin
state and is relayed back to the experiment encoded as a frequency to avoid
having helicity correlated levels in the electronics racks. By running the
full data acquisition system and ion source, with a current source
replacing the main ion chambers, cross talk from the helicity signal was
shown to be negligible.

After the ion source, a Wien filter orients the spin direction so
that it is vertical in the cyclotron.  The 221.3 MeV beam is extracted from
the cyclotron and a solenoid - dipole, solenoid - dipole magnet sequence
rotates the spin to the longitudinal direction. The magnets can be set to
rotate spin-up in the cyclotron to either right handed ($+$) or left
handed ($-$) helicity at the parity apparatus.

In the last section of beam line, the longitudinally polarized beam passes
first through a series of diagnostic devices -- a set of two beam intensity
profile monitors (IPMs) (reference \cite{Berd91} describes an older
version) and a pair of transverse polarization profile monitors (PPMs)
\cite{Berd01}. A third IPM is located immediately in front of the $LH_2$
target, inside the cryostat.

The $LH_2$ target has a flask of 0.10 m diameter and a length of 0.40 m
with flat and parallel end windows. Rapid (5 L/s) circulation of the
$LH_2$ reduces the density gradients to an acceptable level.

The TRICs contain field shaping electrodes plus guard rings to ensure a
uniform sense region, 0.15 m wide by 0.15 m high by 0.60 m long between the
parallel electrodes. The TRICs are filled with ultra-high purity hydrogen
gas and operated at a pressure of about 150 Torr and a high voltage of -8
kV. The entrance and exit windows are located approximately 0.9 m from the
center to range out spallation products and thus prevent their entering the
active region. False parity violating signals due to the buildup of
radioactivity as a result of beam exposure were estimated to be negligible.

The sensitivity to each kind of helicity correlated modulation is measured
in separate calibration runs in which each beam parameter (intensity, size,
position, transverse polarization) is intentionally modulated at an
enhanced level. The sensitivity is then multiplied by the coherent
modulation measured during actual data taking and a correction is applied.

Beam position information is derived from the IPM signals, based on
secondary emission from thin nickel foil strips placed between thin
aluminum foils. The 31 harp signals from each of six planes are
individually amplified and digitized to provide the beam intensity profiles
in each spin state. The beam centroids at two IPM locations are obtained
through integration of the discrete distributions; a corresponding
correction signal is used to drive feedback loops to a pair of horizontal
and vertical fast, ferrite-cored steering magnets. This position servo
holds the beam to within $50 \mu m$ of the ``neutral axis'' on which the
experiment is insensitive to average transverse polarization components.
Sensitivities to helicity-correlated position and size modulations are
determined with the beam unpolarized and enhanced modulations introduced
using fast, ferrite cored magnets synchronized to the spin sequence.

Transverse polarization components are measured by the PPMs, which are
based on p-p elastic scattering using $CH_2$ blade targets that are scanned
through the beam at the beginning of each spin state, immediately prior to
the TRIC integration interval. Each PPM contains detector assemblies for
`left', `right',`down', and `up' scattered protons. The $CH_2$ blades are
1.6 mm transverse to the incident beam direction and 5.0 mm along the beam
direction and move through the beam on a circle of 0.215 m radius with a
frequency of 5 revolutions per second. Each PPM has four blades, two which
scan the polarization profile in the horizontal direction and two which
scan the polarization profile in the vertical direction. By scanning
transversely polarized beams horizontally and vertically, the polarization
``neutral axis'' mentioned above can be found.

To extract $A_z$, measurements in alternating beam helicity are acquired
according to an eight state spin cycle. A switching pattern based on the
sequence ($+--+-++-$) or its complement, is used, canceling both linear and
quadratic drifts. Each spin state lasts 25 ms, comprising settling time,
time for one polarization scan, and a 16.67 ms (1/60 s) TRIC and IPM
integration window. The master clock for sequencing the whole experiment,
including helicity changes, is derived from optical encoders mounted on the
PPM drive shafts. All eight blades of the two synchronized PPMs pass once
through the beam in one 200 ms cycle. The 1/60 s integration time is chosen
to reject all noise at 60 Hz and its harmonics. To permit a check to be
made for unrejected 60 Hz noise, a small controlled phase slip is
introduced so that the master clock drifts through one complete cycle of
the 60 Hz line every 18 minutes.

The parity detection apparatus attains minimal sensitivity to beam current
modulations by precision analog subtraction of the ionization current
signals of the two TRICs. To tune the subtraction for minimum sensitivity
to coherent intensity modulation, an artificially enhanced ($\sim 0.1$\%)
helicity-correlated current modulation can be introduced. Under real data
taking conditions, helicity-correlated current modulation $\Delta I / I =
(I^+ - I^-)/(I^+ + I^-)$ does not normally exceed  a few parts in $10^5$.
Periods of enhanced coherent current modulation are also interleaved with
the ``real'' data so that corrections for $\Delta I / I$ can be made.

\begin{figure}[h]
\includegraphics[width=\linewidth]{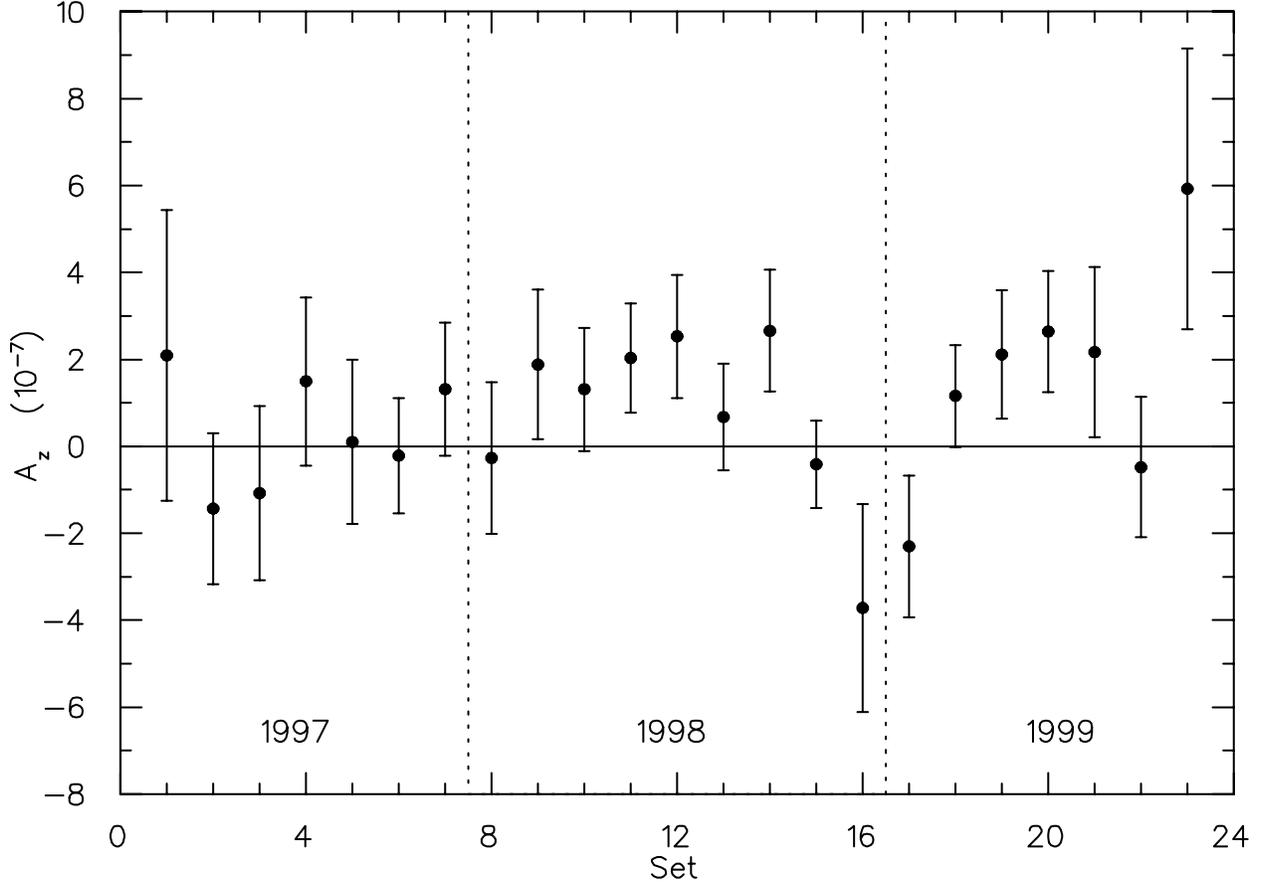}
\caption{Corrected $A_z$ data for each of the 23 data sets.}
\label{data}
\end{figure}

Because the proton energy is on average 27 MeV lower in TRIC2 than in
TRIC1, a small change in incident energy changes the signal from TRIC2 more
than the signal from TRIC1, and coherent energy modulation appears as a
false $A_z$ signal. The sensitivity to coherent energy modulation was
determined using a RF accelerating cavity placed upstream of IPM1 in the
beam line. The measured sensitivity of $(2.9 \pm 0.3) \times 10^{-8}
eV^{-1}$ is in excellent agreement with predictions based on the variation
of stopping power with energy. Helicity-correlated extracted energy
modulations cannot be directly measured at the parity apparatus, so an
appropriate average of data taken with the two beam line helicity tunes is
used to minimize the effect on $A_z$. In addition, interleaved measurements
of intrinsic injected energy modulation at OPPIS together with measurements
at the parity apparatus of sensitivity to enhanced energy modulation give
independent upper limits to the size of the resulting false asymmetries.

\begin{table}
\caption{Summary of helicity correlated beam properties. The table
shows the average value of the coherent modulation, the net correction made
for this modulation, and the uncertainty resulting from applying the
correction.}
\begin{tabular}{lcr}
\toprule
 & & \\
 Property & Average Value & $10^7 \Delta A_z$  \\
 & & \\
\colrule
 $A_z^{uncorrected} (10^{-7})$ & $1.68 \pm 0.29(stat.)$ & \\
 $y*P_x (\mu m)$ & $-0.1 \pm 0.0$ & $-0.01 \pm 0.01$ \\
 $x*P_y (\mu m)$ & $-0.1 \pm 0.0$ & $0.01 \pm 0.03$ \\
 $\langle yP_x \rangle (\mu m)$ & $1.1 \pm 0.4$ & $0.11 \pm 0.01$ \\
 $\langle xP_y \rangle (\mu m)$ & $-2.1 \pm 0.4$ & $0.54 \pm 0.06$ \\
 $\Delta I/I (ppm)$ & $15 \pm 1$ & $0.19 \pm 0.02$ \\
 $position + size$ &           & $     0  \pm 0.10$ \\
 $\Delta E(meV at\, OPPIS)$&   7--15      & $  0.0  \pm 0.12$ \\
 Total & & $0.84 \pm 0.17 (syst.)$ \\
\hline
 & & \\
 $A_z^{corr} (10^{-7})$ &
\multicolumn{2}{l}{$0.84 \pm 0.29(stat.) \pm 0.17(syst.) $} \\
 $\chi_{\nu}^2 (23 sets)$ & 1.08 & \\
 & & \\
\botrule
\end{tabular}
\label{table1}
\end{table}

The data were taken in three one month-long running periods, divided into
23 sets of alternating beam line helicity. Each data set contains a series
of runs, as well as a large number of control measurements, including
polarization neutral axis scans, and measurements of the sensitivity to
various helicity correlated modulations. Helicity correlated modulations in
beam position ($\Delta x$,$\Delta y$), beam size ($\Delta \sigma_x$,$\Delta
\sigma_y$), beam intensity ($\Delta I/I$), transverse polarization coupled
to beam position ($y*P_x$,$x*P_y$), first moments of transverse
polarization ($\langle yP_x \rangle$,$\langle xP_y \rangle$) and energy
modulation at the ion source were considered. Figure \ref{data} shows the
corrected data for the 23 data sets. The uncertainties shown include all
contributions except energy modulation. Table \ref{table1} summarizes the
helicity correlated beam properties, the net correction made for each, and
the uncertainty in each correction. The uncorrected $A_z$ of $(1.68 \pm
0.29)\times 10^{-7}$ becomes $(0.84\pm 0.29(stat.) \pm 0.17(syst.)) \times
10^{-7}$. The reduced $\chi_{\nu}^2 = 1.08$ shown in Table 1 is derived
from the 23 corrected asymmetry values assuming $A_z$ is a constant and
that the uncertainties are as shown in Figure \ref{data} (i.e. not
including uncertainty in the energy modulation correction). Uncorrected
false $A_z$ from energy modulation will have opposite sign in opposite
beamline helicities and increase the $\chi_{\nu}^2$. That the
$\chi_{\nu}^2$ is $1.08$ supports the assertion that contributions from
uncorrected energy modulation effects are small. An overall systematic
uncertainty of $\pm 0.12$ has been assigned for energy modulation.

Note that $A_z$ was measured in a finite geometry with a thick target and
requires a further multiplicative correction of $1.02 \pm 0.02$ when
comparing to theoretical predictions at the $(^1S_0 -^{3\!\!}P_0)$ zero
crossing energy (225 MeV) and integrated over all angles.

The parity violating longitudinal analyzing power in $\vec{p}p$ scattering
has already attracted considerable theoretical interest, and many
calculations have been made \cite{Dris89,Dris90,Icqb94,Grach93}. One of the
main sources of uncertainty in these calculations is the value of the weak
meson nucleon couplings which, up to now, were not sufficiently constrained
by experiment.

The relations between $A_z$ and the the weak couplings have been recently
recalculated by Carlson, {\em et al.} \cite{Carl01} using the Argonne
$v_{18}$ (AV-18) potential \cite{Wirin95}, the Bonn 2000 (CD-Bonn)
\cite{Mach01} strong interaction coupling constants, and including all
partial waves up to J=8. Figure 3 shows the resultant limits on the weak
meson nucleon couplings $h^{pp}_\rho$ and $h^{pp}_\omega$ imposed by the
low energy results \cite{Evers,Kist87} and the present TRIUMF result. The
$\pm 1 \sigma$ bands are based on quadrature sums of the statistical and
systematic errors. Also shown are the DDH ``best guess'' and ``reasonable
range'' \cite{DDH80}.

\begin{figure}[t]
\includegraphics[width=\linewidth]{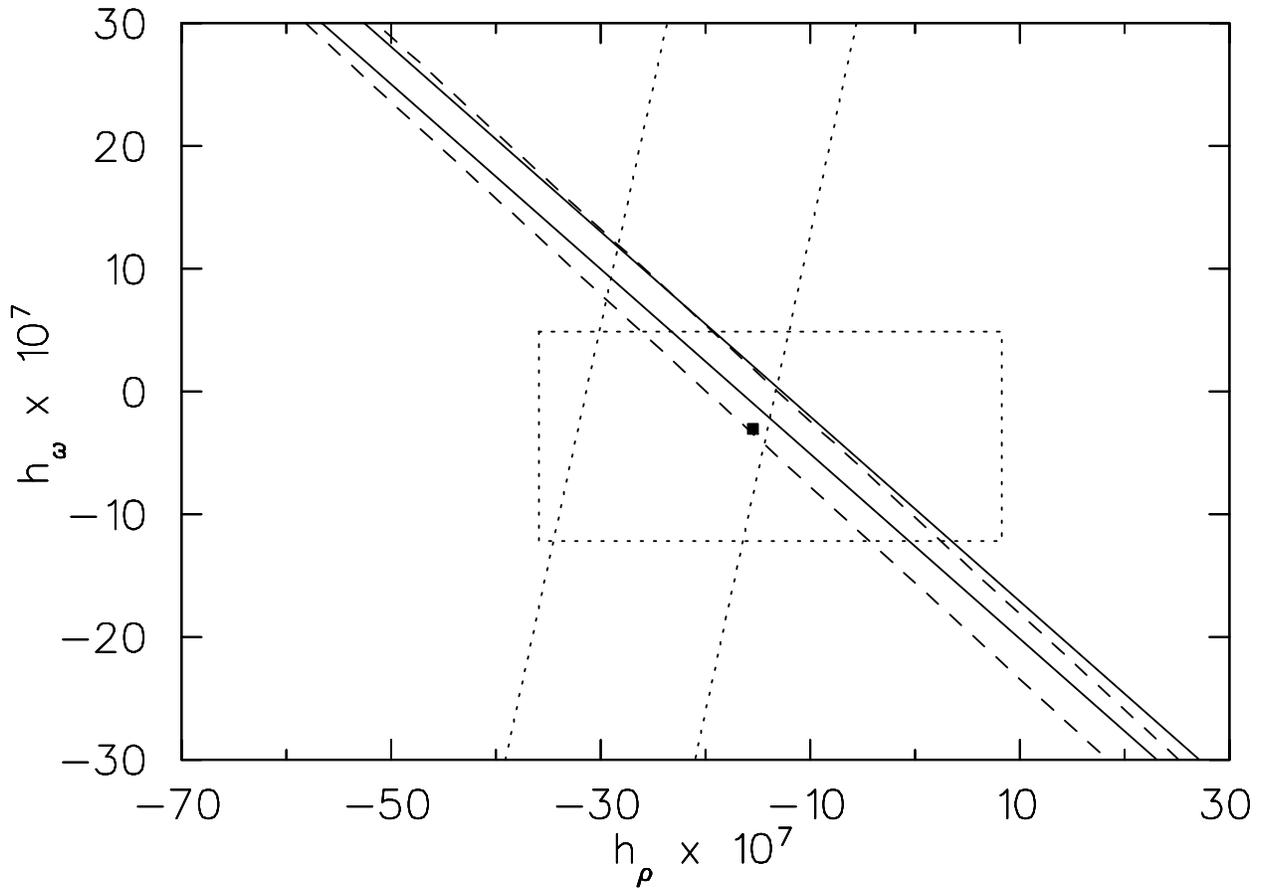}
\caption{Current constraints on the weak meson nucleon couplings based on
experimental data and recent calculations\cite{Carl01}. The bands are the
constraints imposed by different experiments (Bonn 13.6 MeV, dashed; PSI 45
MeV, solid; TRIUMF 221 MeV, dotted). The filled square and dotted rectangle
are the DDH ``best guess'' and ``reasonable range'' respectively. Couplings
are in units of $10^{-7}$.}
\label{couplings}
\end{figure}

Although the experimental error of the present experiment is comparable to
those at the lower energies, the relation between $A_z$ and the weak
couplings is such that the width of the error band in the $h^{pp}_\rho$
and $h^{pp}_\omega$ plane is correspondingly larger than those from the low
energy measurements. A new experiment is being planned at TRIUMF to improve
this situation by significantly reducing the error on $A_z$. Nevertheless,
the present TRIUMF measurement has already greatly reduced the acceptable
range of {\em both} $h^{pp}_\rho$ and $h^{pp}_\omega$ over what was the
case when only the low energy measurements were available.

This work was supported in part by the Natural Sciences and Engineering
Research Council of Canada; TRIUMF receives federal funding via a
contribution agreement through the National Research Council of Canada.

\end{document}